\documentstyle[preprint,aps]{revtex}

\begin{document}

\draft

\title{Gravitational waves from rapidly
rotating white dwarfs}

\author{William A.\ Hiscock\footnote{e-mail: hiscock@montana.edu}}

\address{Department of Physics, Montana State University, Bozeman,
Montana 59717-3840}

\date{July 22, 1998}
\preprint{MSUPHY98.16}
\maketitle

\begin{abstract}

Rapidly rotating white dwarfs in cataclysmic variable systems may be
emitting gravitational radiation due to the recently discovered
relativistic r-mode instability.  Assuming that the four most rapidly
rotating known systems are limited in rotation rate by the
instability, the amplitude of the emitted gravitational waves is
determined at Earth for both known rapid rotators and for a model
background caused by a galactic population of such systems.  The
proposed LISA and OMEGA space-based interferometer gravitational wave
detectors could observe such signals and determine whether the r-mode
instability plays a significant role in white dwarf systems.

\end{abstract}
\newpage

Andersson \cite{and1} has recently discovered a new relativistic
instability in rapidly rotating stars.  All rotating stars formed of
perfect fluid possess unstable nonaxisymmetric modes (``r-modes'')
which will emit gravitational radiation, carrying away angular
momentum and slowing the rotation of the star \cite{FM98}.
Whether these modes are important dynamically in a realistic star
depends on the viscosity, which tends to damp these modes, as is
the case in the previously studied f-modes \cite{L95,LH82}.
For any given stellar model, there is a critical rotation rate
which may be defined, such that the star is stable due to viscous
damping when rotating below this rate, but unstable above.
Estimates of this critical period for neutron stars show that
the r-mode instability may be of importance in the
evolution of newly born hot, rapidly rotating
neutron stars, spinning them down to the critical rate
\cite{LOM98,AKS98}. The resulting gravitational radiation may be an
interesting source for ground-based interferometers, such as
LIGO and GEO600 \cite{Owen98}.

Recently, Andersson, Kokkotas, and Stergioulas (hereafter AKS)
\cite{AKS} have
considered whether the r-mode instability might play a role in
rapidly rotating white dwarfs, particularly the
DQ Herculis stars. These stars are magnetized cataclysmic
variables (CVs) with rotation periods as low as ~ 30 s \cite{P94}.
They accrete material from a low-mass main sequence companion
in a binary system. A simple estimate of the critical rotation period
made by AKS is intriguingly close to the actual rotation periods of
the more rapidly rotating DQ Her stars. If the r-mode instability
is active in such a system, then it represents a steady-state source
of gravitational radiation. The star would be rotating at the
critical limit, where all excess angular momentum accreted would be
emitted as gravitational radiation. The equivalent situation for
an f-mode limited neutron star was previously considered by
Wagoner \cite{Wag84}. AKS also found that the growth
timescale of the unstable r-mode is so great in a white dwarf
(of order $10^8$ yr) that it appears questionable whether the
mode could play a defining role in the system's dynamics.
Nevertheless, there is at present considerable
uncertainty in the analysis of the r-mode instability in
white dwarfs, and the DQ Her stars are complex systems. If
the most rapidly rotating DQ Her white dwarfs are potentially
rotating at the critical limit, it is worth considering 
whether the emitted gravitational waves (GW) are potentially 
detectable.

The purpose of this paper is to show that if the r-mode 
instability is actually excited in the rapidly rotating DQ
Her stars, then the gravitational radiation emitted is
potentially detectable. Gravitational
wave astronomy may then, in principle, be used to constrain or 
determine the relevance of the r-mode instability to 
these systems.  The gravitational wave frequency of
the dominant r-mode, with $ l = m = 2$, is $ f = 4/(3P)$, where
$P$ is the rotation period of the star. For the most rapidly
rotating white dwarfs, this means the emitted gravitational
waves will be in the frequency band of $10^{-2} - 10^{-1}$
Hz, the domain of the proposed space-based laser interferometer
detectors, such as LISA \cite{lisa} and OMEGA \cite{omega}.

The critical rotation period for a white dwarf of mass $M$,
radius $R$, and temperature $T$, was estimated by AKS to be
\begin{equation}
        P_c \simeq 27 \left( {M \over M_\odot} \right)^{1/24}
        \left( {R \over 0.01R_\odot} \right)^{11/8} \left(
        { T \over 10^5 K } \right)^{1/24} {\rm } \;\; s.
\label{pcrit}
\end{equation}
There are four candidate stars with short rotation periods which
are within a reasonable range of this estimated critical period.
These are: WZ Sge, with a period of 28 s, AE
Aqr with period 33 s, V533 Her at 63.6 s, and DQ Her at 71.1 s.
For the purposes of this paper, all four of these stars will be
assumed to be rotating at the critical limit, and hence be
emitting GW due to the r-mode instability. The remaining known
members of the DQ Her family have
considerably longer rotation periods, from 206 s to 7188 s
\cite{P94}, and will be ignored here. 

The four most rapidly rotating DQ Her stars may be treated as
periodic sources of gravitational radiation, since their 
rotation frequencies are known {\it a priori} with precision
and are essentially constant. These four systems are also used to
develop plausible values for the spatial density and gravitational
wave luminosity of rapidly rotating white dwarfs throughout 
the Galaxy. These values are then used to determine the 
cumulative background of gravitational radiation due to 
the superposed signals of the rapidly rotating DQ Her type 
systems throughout the Galaxy.

The stellar systems of interest are assumed to be rotation
limited by the emission of gravitational radiation associated
with the r-mode instability. All excess angular momentum
carried by the matter accreted from the companion star is
converted into gravitational radiation by the instability.
Following Wagoner \cite{Wag84}, this implies that
\begin{equation}
        {dJ \over dt} \simeq (M R)^{1/2} \, {\dot M} -
        {\dot J}_{GR} = 0 \;\; ,
\label{balance}
\end{equation}
where $J$ is the total angular momentum of the accreting
dwarf, and ${\dot J}_{GR}$ represents the loss of angular
momentum to gravitational radiation. Units have been chosen
so that $ G = c = 1$. The angular momentum carried off in the
gravitational radiation may be related to the energy lost, and
to the dimensionless amplitude of the gravitational wave, $h$,
as observed at Earth, by
\begin{equation}
        {\dot J} = - {m \over \omega} {\dot E} \;\; ,
\label{ejdot}
\end{equation}
\begin{equation}
        h^2 = \left({2 \over {\omega r}} \right)^2 \vert {\dot E}
        \vert \;\; ,
\label{hdef}
\end{equation}
where $m$ and $\omega$ are respectively the azimuthal index and
frequency of the mode, and $r$ is the distance from
the source to the Earth.
Combining Eqs(\ref{balance}-\ref{hdef}) then yields
\begin{equation}
        h = { {2 \, (M R)^{1/4} \, {\dot M}^{1/2}} \over
        { m^{1/2} \,\omega^{1/2} \, r}} \;\; ,
\label{h}
\end{equation}
or, for the dominant mode, assuming the star is rotating at the
critical period,
\begin{equation}
        h = 1.88 \times 10^{-24} \left( {P_c \over 30 {\rm s}}
        \right)^{1/2} \left({ M \over M_\odot} \right)^{1/4}
        \left({ R \over 0.01 R_\odot} \right)^{1/4}
        \left({ {\dot M} \over 10^{-10} M_\odot /{\rm yr}} \right)
        ^{1/2} \left({{\rm kpc} \over r} \right) \; .
\label{hnum}
\end{equation}
If the four rapidly spinning white dwarfs are assumed to be
rotating at $P_c$, so that they are balanced between accretion
of angular momentum and emission via gravitational radiation, then
the amplitude of the wave at Earth may be calculated using
Eq.(\ref{hnum}). The pertinent properties of the four
stellar systems under consideration, and their resulting
values of $h$, are listed in Table 1. Values for the spin period,
mass, accretion rate, and distance were taken from
Refs. \cite{AKS,P94,Warn76}; the radii were then determined
by assuming the stars are CO white dwarfs \cite{ST}. No
mass value for V533 Her was found in the literature, so a value
of $1 M_\odot$ was assumed. The gravitational wave amplitudes for
these nearby ( $r < 1 \; {\rm kpc}$) systems are found to be in
the range $h \sim 10^{-23} - 10^{-22}$.

Since the four known rapidly rotating white dwarfs are all
nearby, it is reasonable to assume that these represent only the
nearest (and brightest, electromagnetically) such stars. A
population of such stars throughout the galaxy will create a
background of gravitational waves which may be
detectable. In order to determine the GW spectrum of such a
background the properties of the known stars must be used
to model a galactic population. 

Since the properties of the four known nearby systems
are the only data available to define the model of 
the galactic distribution of such rapidly rotating
white dwarfs, the result should be recognized as being at
best a crude approximation. The four known rapid rotators
fall naturally into two groups of two. The
two nearby systems, WZ Sge and AE Aqr, are quite rapidly rotating,
with similar periods and similar (low) mass accretion
rates, and hence presumably lower gravitational wave luminosity.
The two more distant systems, V533 Her and DQ Her, are
more slowly rotating, again have similar periods, and have higher 
accretion rates. It thus seems sensible to construct a model in
which the luminosity and galactic density of systems depends on the
rotation period. The model distribution will assume that all DQ Her
stars with periods between $20 {\rm s}$ and $80 {\rm s}$ are
rotating at their critical periods and emitting GW due to the
r-mode instability.

The gravitational wave amplitude associated with the four known
systems is well fit by the power law:
\begin{equation}
      h \, = \, 2.0 \times 10^{-28} \left({ f \over {\rm Hz}}
      \right)^{-3} \; \left({{\rm kpc} \over d} \right) \; \; .
\label{hmodel}
\end{equation}

The spatial density of rapid rotators as a function of period will
be modeled by noting that the two rapid rotators (WZ Sge and AE Aqr)
are within 100 pc; while the slower DQ Her and V533 Her are within
1 kpc. A density function is then chosen for the solar neighborhood
which is a power law of rotation period (or GW frequency, $f$),
normalized to yield a few fast rotators within 100 pc and a few
slower rotators within 1 kpc,
\begin{equation}
      {d\rho_\odot \over dP} = 5 \times 10^{-7} \left({P \over
      30 {\rm s}}\right)^{-6} \; {\rm pc}^{-3}\; {\rm s}^{-1} \;\; .
\label{rhosol}
\end{equation}
This choice of distribution implies that there should be numerous
fast ($P \sim 30 {\rm s}$) rotators between distances of 100 pc
and 1 kpc which have not yet been discovered. 

The distribution of rapidly rotating white dwarfs in the galaxy is
assumed to follow the galactic disk population of CVs, so that
\begin{equation}
      {d\rho \over dP} = {d\rho_0 \over dP} \, \exp(-R/R_0)\,
      \exp(-\vert z \vert /z_0) \; \; ,
\label{gal}
\end{equation}
in standard Galactocentric coordinates, with $R_0 = 3.5 \; {\rm kpc}$,
$z_0 = 120 \; {\rm pc}$, and the central density, $d\rho_0/dP$ is given
by $d\rho_0/dP = d\rho_\odot/dP\, \exp(R_\odot / R_0)$ \cite{HBW}.
The solar Galactocentric radius is taken to be $R_\odot = 8.5
\;{\rm kpc}$.

The proposed space-based interferometers, LISA and OMEGA, will
integrate the signal from persistent sources for a characteristic
time of $ \sim 4 \; {\rm months}$, yielding a bandwidth of $10^{-7}
\; {\rm Hz}$. The rotating white dwarf sources are assumed to have
periods of $20-80\; {\rm s}$, covering a range in frequency space from
$ 1.67 \times 10^{-2} - 6.67 \times 10^{-2}\; {\rm Hz}$, which will
contain $5 \times 10^5$ frequency bins of width $df = 10^{-7}
\; {\rm Hz}$.
The number of sources per bin can be determined by integrating
Eq.(\ref{gal}) over the galaxy, converting from period to
frequency dependence, and multiplying by the width of the frequency
bin, to find
\begin{equation}
      {dN \over df}\,df = 2.4 \times 10^{5} \left({ f \over
      {\rm Hz}} \right)^4 \; \; .
\label{nbin}
\end{equation}
Over the relevant range of frequencies the number of sources 
per bin then varies
from 0.02 to 5; over the majority of the frequency domain, most
bins are empty of any source. Since there are not large numbers of
sources per bin, the background is not stochastic; after 4 months
of integration most nonempty bins will be resolved into individual
periodic sources.

A simulated GW spectrum from a galactic population of r-mode
unstable white dwarfs which could be potentially detected by LISA
or OMEGA with a 4 month integration is illustrated in Figure (\ref{fig1}).
This simulated spectrum was constructed as follows. First,
the number of sources in each of the $1/2$ million
bins was determined from Eq.(\ref{nbin}); if this number 
is one or more, then the sources in the
bin are placed randomly in the Galaxy, weighted by the distribution
function of Eq.({\ref{gal}), using
a Monte Carlo routine. The GW spectral amplitude is then calculated using
Eq.(\ref{hmodel}) and the bandwidth. Finally, if there is more than
one source in a bin, their amplitudes are added incoherently
(square root of the sum of squared amplitudes). The four
known rapidly rotating white dwarfs are indicated by the 
labeled points. It is important to emphasize again that most
of the bins, particularly at the lower frequencies, are actually
empty; this is not apparent due to the thickness of the ink lines
in the figure.

The figure also illustrates the root spectral density of the
instrument noise for
the proposed LISA and OMEGA gravitational wave detectors. These
proposed systems each use several spacecraft arranged in an
equilateral triangle, whose sides form laser interferometers. Both
proposed missions plan to use 1 watt lasers, 30 cm optics, and
achieve a drag-free performance of order $10^{-15} \;
{\rm m}\;{\rm s}^{-2}\;{\rm Hz}^{-1/2}$. Their sensitivity to
gravitational waves is determined by these three parameters
and the size of the interferometer arms. The
LISA plan calls for a heliocentric orbit with an arm baseline
of $5 \times 10^9 \; {\rm m}$; OMEGA is intended to be geocentric,
with a shorter arm length, only $ 10^9 \; {\rm m}$. The longer
baseline would make LISA more sensitive at low frequencies,
while OMEGA's performance would be better at higher frequencies.
The sensitivity curve shown for LISA does not agree with that
often displayed (see, e.g., Ref.\cite{lisa}, page 20) at frequencies 
above $10^{-2} \; {\rm Hz}$. This is because the degradation
in sensitivity at high frequencies has usually been assumed to 
begin at $f = (2 \ell)^{-1}$, where $\ell$ is the interferometer
baseline. However, careful analysis of the transfer function
\cite{Hellings83,Schilling97} shows that for any space
interferometer, the degradation in sensitivity actually 
begins at a lower frequency, namely $ f = (2\pi\ell)^{-1}$.
This lower onset frequency has been used consistently for
both instruments, so that their relative sensitivity is correctly
portrayed in Fig.(\ref{fig1}).

Both the galactic background and the periodic signals of nearby
sources are seen in Figure (\ref{fig1}) to be within the 
detection limits of either interferometer system.  If the
source model is made more conservative, by assuming that
only the most rapid rotators,
with periods closest to the estimate of \cite{AKS}, 
possess active r-mode instabilities (e.g., $ P < 35$ s, or $f > 3.8
\times 10^{-2} \; {\rm Hz}$), then OMEGA could still detect
a large number of sources, while LISA would be limited to only
the nearest and brightest. 

In summary, if the r-mode relativistic instability is active
in rapidly rotating white dwarfs, then the gravitational waves
emitted would be detectable by proposed space-based interferometer
systems. The actual relevance of the r-mode instability to rapidly
rotating white dwarfs could then be determined observationally.
As these sources are higher frequency than most other sources
considered for the space-based interferometers, shorter baseline
instruments would better suited to their detection.

This research was supported in part by National Science Foundation
Grant No. PHY-9734834. The author wishes to thank R. Hellings and
S. Larson for helpful discussions.

\begin{figure} 
\caption{The instrument noise curves for LISA and OMEGA are
compared to the estimated spectral amplitude for r-mode limited
rapidly rotating white dwarfs. A bandwidth
of $10^{-7} {\rm Hz}$ is assumed, which allows the majority of
stars be resolved into individual GW spectral lines. 
The four known rapid rotators, indicated by the labeled
points, are, from left to right,  DQ Her, V533 Her, AE Aqr, WZ Sge.} 
\label{fig1} 
\end{figure}

\begin{table}
\caption{The properties of the four rapidly rotating DQ Her
systems are given, along with the values of $h$ which follow
by assuming their rotation rate is limited by the r-mode
instability.}
\begin{tabular}{ccccccc} 
System&Spin period (s)&${M \over M_\odot}$&${R \over 0.01R_\odot}$
&${{\dot M} \over 10^{-10} M_\odot/{\rm yr}}$&$ r({\rm pc})$&$h$\\
\tableline
WZ Sge&$28$&$1.4$&$0.2$&$1.6$&$75$&$2.27 \times 10^{-23}$\\
AE Aqr&$33.1$&$1.25$&$0.5$&$2.0$&$90$&$2.76 \times 10^{-23}$\\
V533 Her&$63.3$&$1.0$&$0.8$&$63.$&$1000$&$2.06 \times 10^{-23}$\\
DQ Her&$71.1$&$0.83$&$1.0$&$126.$&$420$&$7.38 \times 10^{-23}$\\
\end{tabular}
\label{table1} 
\end{table}

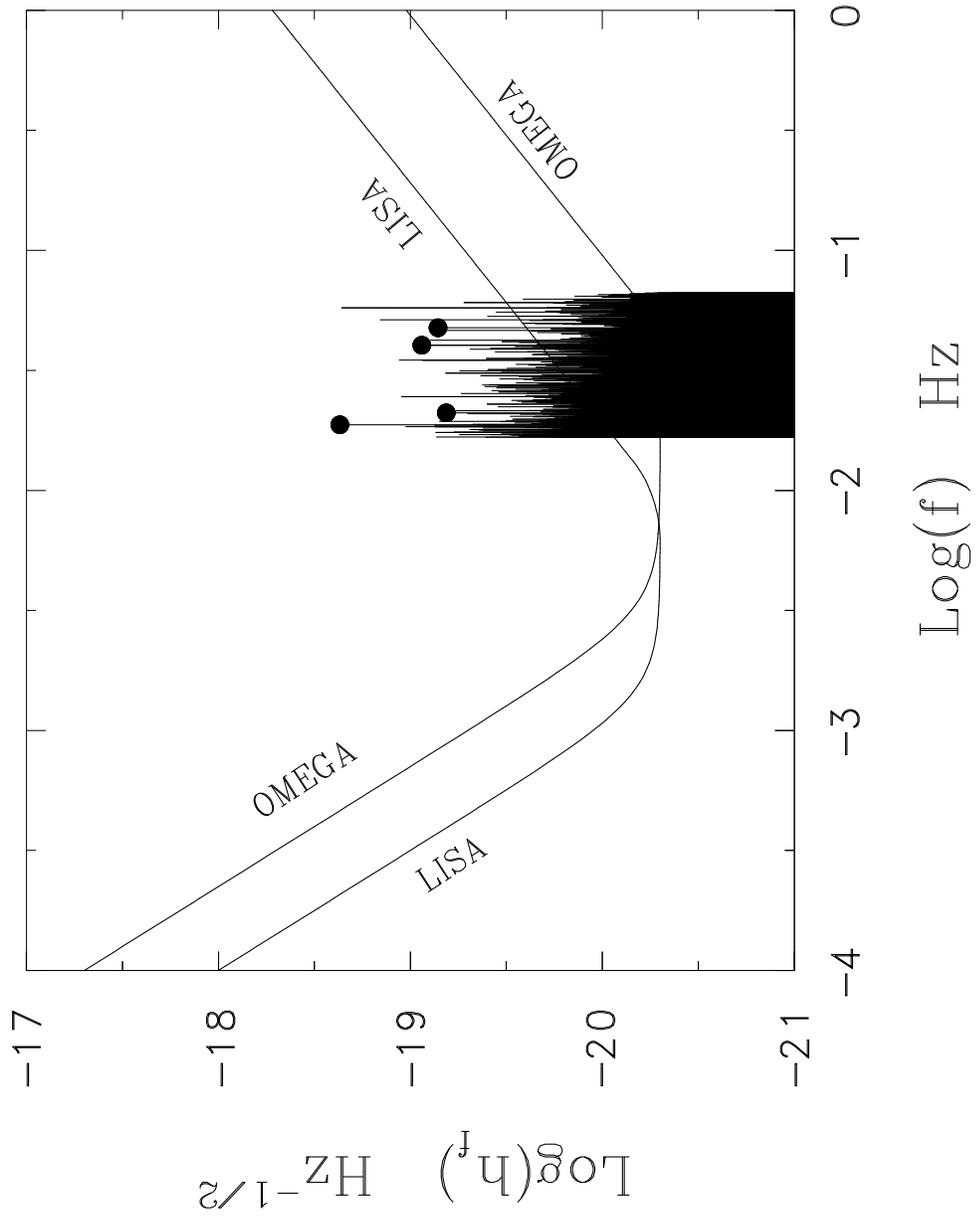
\end{document}